\documentclass{PoS}
\usepackage{amsmath}
\usepackage{mathrsfs}

\newcommand{\dHybridR}{{\tt dHybridR}}
\newcommand{\citep}{\cite}

\newcommand{\vsh}{V_{\rm sh}}

\title{The Issue with Diffusive Shock Acceleration}

\ShortTitle{The Issue with  DSA}

\author{\speaker{Damiano Caprioli}\\
         University of Chicago, 5640 S Ellis Ave., Chicago, IL 60637 (USA)\\
        E-mail: \email{caprioli@uchicago.edu}}

\author{Colby Haggerty\\
        University of Chicago, 5640 S Ellis Ave., Chicago, IL 60637 (USA)\\
        E-mail: \email{chaggerty@uchicago.edu}}

\abstract{We discuss the recent developments in the theory of diffusive shock acceleration (DSA) by using both first-principle kinetic plasma simulations and analytical theory based on the solution of the convection/diffusion equation. 
In particular, we show how simulations reveal that the spectra of accelerated particles are significantly steeper than the $E^{-2}$ predicted by the standard theory of DSA for strong shocks, in agreement with several observational pieces of evidence.
We single out which standard assumptions of test-particle and non-linear DSA are violated in the presence of strong (self-generated) magnetic turbulence and put forward a novel theory in better agreement with the particle spectra inferred with multi-wavelength observations of young SN remnants, radio-SNe, and Galactic cosmic rays.
}

\FullConference{36th International Cosmic Ray Conference -ICRC2019-\\
		July 24th - August 1st, 2019\\
		Madison, WI, U.S.A.}

\begin{document}

\section{Introduction} \label{sec:intro}
Diffusive Shock Acceleration (DSA) is a ubiquitous mechanism for producing relativistic particles and non-thermal emission \cite{krymskii77,bell78a,blandford+78,axford+78}.
This special case of first-order Fermi acceleration, which applies to particles diffusing back and forth across the shock discontinuity, is particularly appealing because it naturally produces power-law spectra that depend only on the compression ratio, $r$,  i.e., the ratio of  downstream to upstream gas density. 
For a monoatomic gas with adiabatic index $\gamma=5/3$, the compression ratio and the spectral slope $q$ read:
\begin{equation}\label{eq:qdsa}
    r=\frac{\gamma+1}{\gamma-1+2/M^2} \to 4; \qquad  q_{\rm DSA}= \frac{r+2}{r-1}\to 2, 
\end{equation}
where we  took the limit of Mach number $M\gg1$ (strong shock).
Strictly speaking, spectra are universal  in momentum $p$ and scale as $f(p)\propto p^{-4}$, which means $q=1.5$ for non-relativistic particles; in this work we consider the energy scaling, for simplicity.

When accelerated particles (henceforth Cosmic Rays, CRs) carry a non-negligible fraction of the shock momentum/energy flux, they cannot be regarded as test-particles;
in such \emph{CR-modified shocks} both the shock dynamics and particle spectra deviate from the standard predictions. 
The non-linear theory of DSA (e.g., \cite{jones-ellison91,malkov-drury01} for reviews) suggests that the acceleration efficiency ($\gtrsim 10\%$) required for Supernova Remnants (SNRs) to be the sources of Galactic CRs should induce a shock precursor that leads to a total compression ratio $r\gg 4$, and thereby to CR spectra much flatter than the DSA prediction.
Accounting for the dynamical role of the  CR-driven magnetic turbulence limits such a compression to values $\lesssim 10$ \cite{caprioli+08, caprioli+09a}, but does not alter the theoretical expectation that strong shocks should be efficient accelerators and produce CR spectra with $q<2$.

In this work we outline the tension between the DSA theory and observations (\S\ref{sec:obs}) and the possible solutions that have been suggested  (\S\ref{sec:theory}).
Then, we present some preliminary results  obtained with particle-in-cell (PIC) kinetic simulations of non-relativistic shocks, which provide the first hints that it is possible to have both efficient CR acceleration and spectra appreciably steeper than what predicted by the non-linear DSA theory  (\S\ref{sec:sims}).



\section{The Challenging Observations}\label{sec:obs}
\begin{figure}[tb]
\centering
\includegraphics[trim=3px 3px 3px 3px, clip=true, width=.6\textwidth]{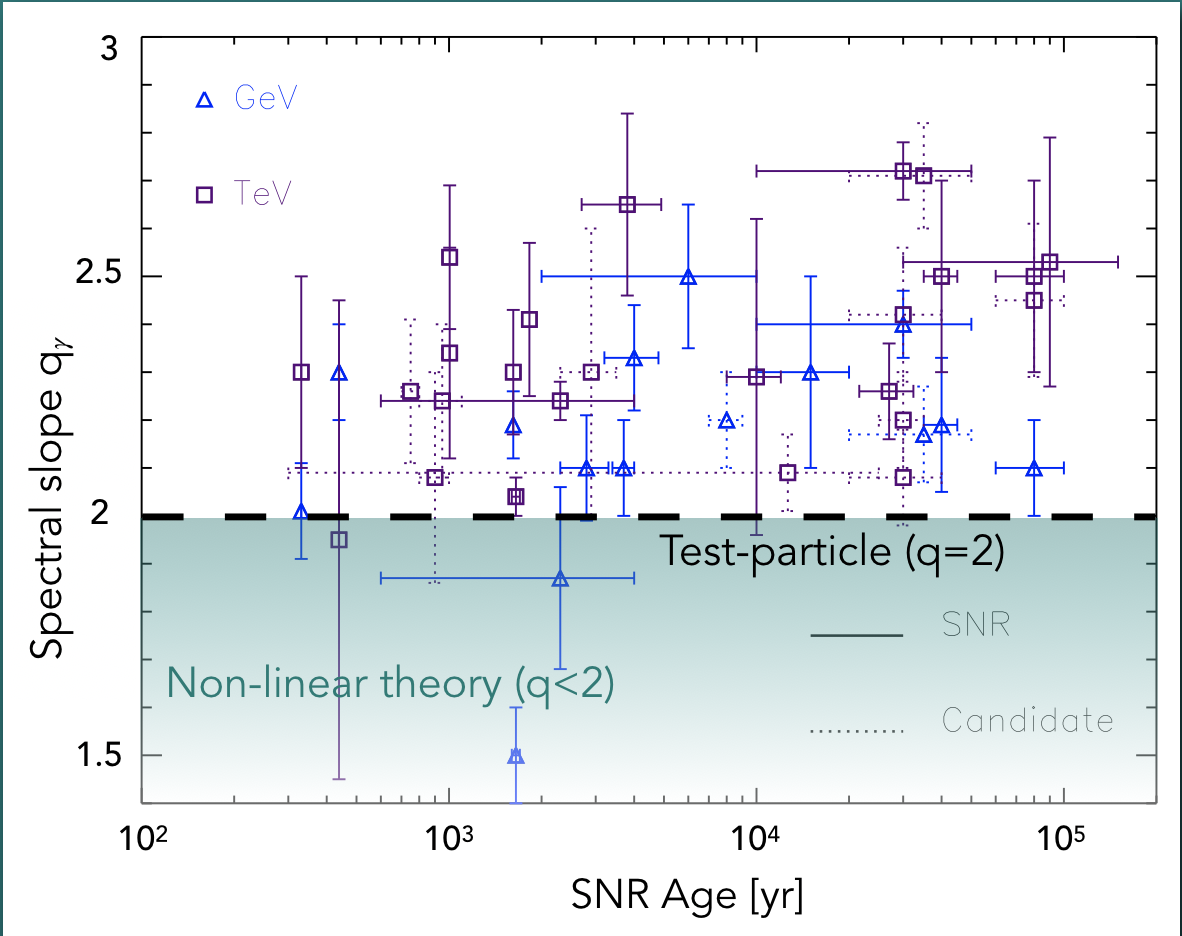}
\caption{Energy spectral indexes $q_{\gamma}$ of the $\gamma$-ray spectra observed in the GeV/TeV bands for SNRs of different ages.
Note that most of the SNRs show spectra steeper than the test-particle DSA prediction and much steeper than what expected when non-linear effects are included. 
Adapted from \cite{caprioli11,caprioli12}.}\label{fig:gamma}
\end{figure}

There are several observational indications that shock acceleration may produce particle spectra steeper than what predicted by the DSA theory.

{\bf $\gamma$-ray emission from SNRs.}
SNRs have been extensively observed by Cherenkov telescopes (HESS, MAGIC, VERITAS, and HAWC) in the TeV energy range and satellites (Fermi and AGILE) in the GeV band.
Very interestingly, the photon spectral index in most of the $\gamma$-ray bright SNRs is inferred to be appreciably larger than 2, typically in the range 2.2--3 \cite{caprioli11, caprioli12, acero+16short},  see Figure \ref{fig:gamma}.
$\gamma$-ray emission may be either  {\it leptonic} (relativistic bremsstrahlung and inverse-Compton scattering, IC) or {\it hadronic} ($\pi^0$ decay);
in the hadronic scenario, the $\gamma$-ray spectrum is parallel to the one of parent hadrons, while IC scattering produces harder photon spectra ($\propto E^{-1.5}$ for an $E^{-2}$ electron spectrum).
Therefore, away from the cut-off of the parent particle distribution, a steep spectrum represents a strong signature of hadron acceleration and suggests that the parent hadrons also have spectrum with $q>2$.
Remarkably, at GeV energies, where synchrotron cooling is not effective, protons and electrons are expected to show the same spectral index \cite{diesing+19,morlino+12}, which means that steep spectra are required even in a leptonic scenario.

{\bf  Radio-SNe.}
Very young SNRs (days to months old), observed in other galaxies in the radio and X-rays, also offer us clues that electrons are typically accelerated to relativistic energies with spectra as steep as $E^{-3}$ \cite{cf06, bell+11}; radiative losses does not seem to be important in this case, either.
Note that these so-called radio-SNe probe a different regime of shock acceleration than Galactic SNRs, the shock velocity still being quite large: $\vsh\gtrsim 10^4$ km s$^{-1}$ and even transrelativistic.

 {\bf Galactic CRs.}
Connected with the problem of accelerating CRs with power-law distributions is the problem of preserving such regular structures during the CR journey from sources to the Earth.
The CR Galactic residence time can be estimated thanks to radioactive clocks such as $^{10}$Be and to the   ratios of secondary to primary species  (e.g.,  B/C), which return the grammage traversed by primary CRs in the Galaxy.
If CRs are produced in the disk and diffusively escape at some distance $H$ ($\sim$\,a few kpc) in the halo, the Galactic residence time is $\tau_{\rm gal}(E)\approx H^2/D_{\rm gal}(E)$, where $D_{\rm gal}(E)$ is the diffusion coefficient that parametrizes CR transport in the Galaxy, assumed homogeneous and isotropic.
The energy dependence of primary/secondary ratios scales as $\tau_{\rm gal}\propto E^{-\delta}$ and is crucial for connecting the spectra injected at sources ($N_{\rm s}\propto E^{-\alpha}$) with those measured at Earth, ($\propto E^{-2.65}$ below the knee \cite{PAMELA14,AMS15}).
The equilibrium CR spectrum can in fact be written as $N_{\rm gal}(E)\propto N_{\rm s}(E) \mathcal{R_{\rm SN}}\tau_{\rm gal}(E)$, which imposes $\delta+\alpha\approx 2.65$.
Since the most recent AMS-02 data constrain $\delta\approx 0.33$  \citep{ams16a}, one finds $\alpha\approx 2.3-2.4$, quite steeper than the DSA prediction for strong shocks.
Note that the cumulative spectrum produced over the SNR history is typically $\sim 0.1$ steeper than the one at the beginning of the Sedov stage, i.e., $\alpha\simeq q+0.1$ \cite{caprioli+10a}.
Despite its simplicity, this diffusive model for CR transport is quite solid because it simultaneously accounts for the measured CR secondary/primary ratios, the diffuse Galactic synchrotron and $\gamma$-ray emission (see, e.g., \cite{ba12a,galprop98,dragon13}), and even the observed anisotropy in the arrival directions of CRs \cite{ba12b}.

\section{The Theoretical Challenge}\label{sec:theory}
The origin of such steep spectra has puzzled theorists, who came up with some possible explanations for their origin.
Again, the challenge here is to allow for shocks to be efficient CR accelerators, which means that the physical process that leads to the spectral steepening has to win against the hydrodynamical effect of the CR pressure that tends to flatten the spectra above a few GeV \cite{jones-ellison91,malkov-drury01}.
A few mechanisms have been proposed \cite{caprioli15p}, but none of them has been verified in self-consistent kinetic simulations of shocks. 

{\bf Magnetic feedback.} 
Recalling that the CRs are actually coupled with the magnetic fluctuations advected with the fluid, rather than with the fluid itself,  changes the effective definition of the compression ratio that enters the DSA slope, as already pointed out in \cite{bell78a}.
In the presence of strong magnetic field amplification \cite{caprioli+14b}, the velocity of the CR scattering centers relative to the fluid may become non-negligible,  and ---provided that waves travel in the ``correct'' direction (away from the shock upstream and/or downstream)--- spectra could become steeper than $E^{-2} $ \cite{zp08b,caprioli11, caprioli12}.
It is intriguing that the magnetic fields inferred in young SNRs may suffice to produce  $q \approx 2.3$ \cite{morlino+12}.

Very recently, a different take on magnetic feedback has been proposed \cite{bell+19};
CR spectra may be steepened because of the energy that they channel into the turbulent magnetic field amplification and the generation of turbulent motions would thereby reduce the net upstream flow, eventually resulting in a decreased compression ratio and a steeper CR spectrum.

\textbf{Neutral return flux.}
Several $\gamma$-ray--bright SNRs expand in partially-neutral media, where the shock dynamics is modified at the zeroth order by  ionization and charge exchange mechanisms  \cite{blasi+12a}. 
In particular, the returning flux of neutrals that undergo charge exchange right downstream of the shock produces a precursor in which the upstream flow is effectively slowed down and heated up, thereby leading to a much weaker shock with a compression ratio $r\ll 4$ and hence $q>2$ \cite{morlino+13}.
Such an effect should be present only for $\vsh\lesssim 3000$\,km\,s$^{-1}$, since for larger $\vsh$ ionization dominates over charge exchange and the neutral return flux vanishes.

\textbf{Trans-relativistic effects.}
The standard DSA theory at non-relativistic shocks hinges on the expansion of the CR transport equation in terms of increasing anisotropy, which generally scale as $\vsh/c$.
Including diffusion parallel to the magnetic field is already an effect $\mathcal O (\vsh/c)$, but very fast shocks and/or oblique shocks (namely that propagate at a large angle with respect to the large-scale magnetic field) may require to include higher-order effects that alter the DSA prediction \cite{bell+11,kirk+96}.

\textbf{Geometry-dependent effects.}
It has been recently suggested \cite{malkov+19} that a spectral steepening may appear also because particles are preferentially injected into DSA at quasi-parallel shocks \cite{caprioli+14a, caprioli+15}; 
in this case a steep CR spectrum would originate from the time convolution of different contributions from quasi-parallel and oblique shock regions, each of them with a specific normalization and maximum energy. 
This effect may disappear when the SNR diameter becomes larger than the coherence length of the background magnetic field, though.

It is rather clear that some of these possibilities are mutually exclusive, for instance when different regimes of $\vsh$ are involved (trans-relativistic and geometry effects may apply to young SNRs, while neutrals require rather slow shocks). 
On the other hand, effects related to magnetic field amplification may work in different fashions during the SNR evolution and need to be validated by a self-consistent theory of CR acceleration including magnetic field generation.

\section{Hybrid Simulations}\label{sec:sims}
\begin{figure}[tb]
\centering
\includegraphics[trim=3px 3px 3px 3px, clip=true, width=.9\textwidth]{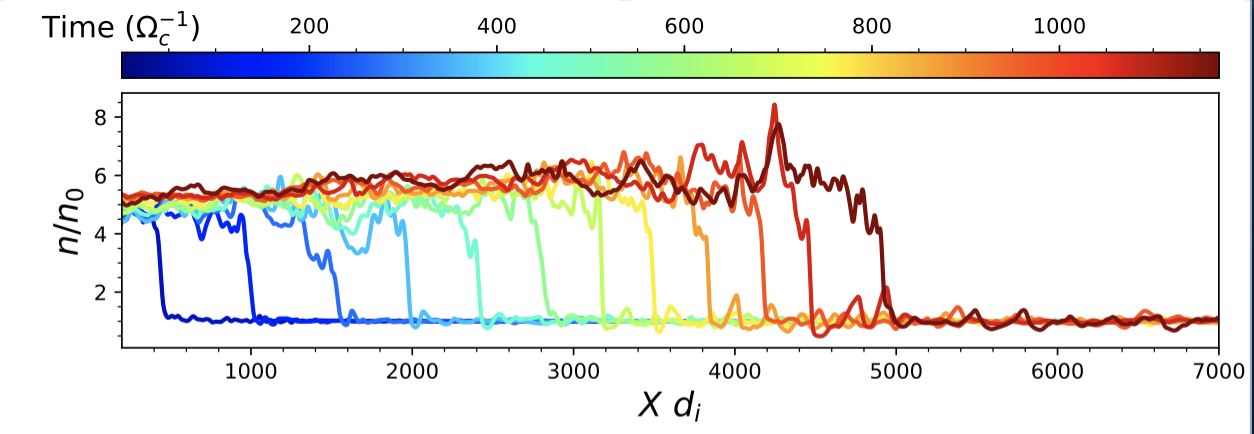}
\includegraphics[trim=3px 3px 3px 3px, clip=true, width=.9\textwidth]{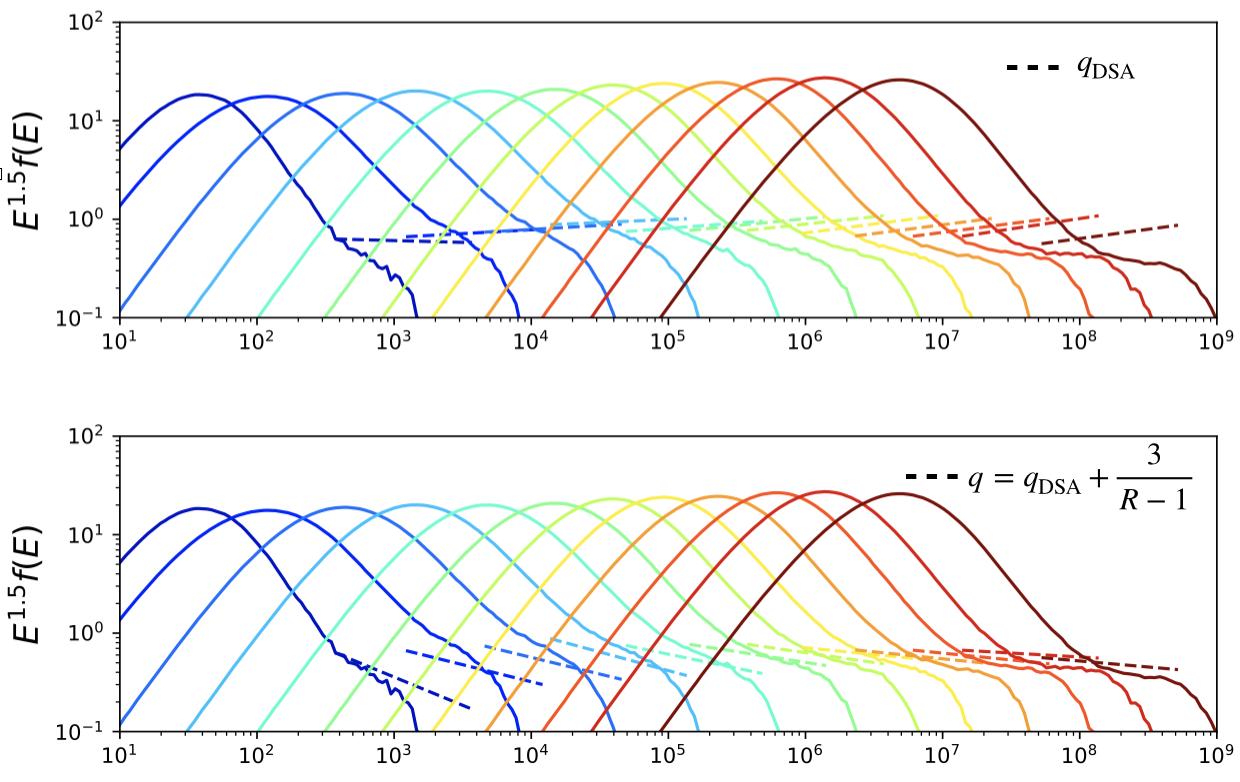}
\caption{\footnotesize 
{\bf Top panel}: Time evolution of the density profile for a strong parallel shock with $M=20$. The post-shock compression ratio increases with time from $R\sim 4$ to $R\gtrsim 7$, a signature of CR-induced shock modification.
{\bf Middle and bottom panels:} Post-shock ion spectra at the same time steps (color coded), shifted on the horizontal axis to match the shock position in the top panel. 
Note how the non-thermal power-law tail gets flatter with time, but always being steeper than the DSA prediction (middle panel). The best fitting for $q$ is given by the formula in the legend of the bottom panel.}
\label{fig:hybrid}
\end{figure}

Kinetic PIC simulations are prominent tools for investigating CR acceleration and the production of magnetic turbulence, especially hybrid simulations, with kinetic ions and neutralizing fluid electrons. 
Hybrid simulations have been used to perform a comprehensive analysis of proton acceleration at collisionless shocks as a function of the strength and topology of the pre-shock magnetic field, the nature of CR-driven instabilities, and the transport of energetic particles in the self-generated magnetic turbulence \cite{caprioli+14a,caprioli+14b,caprioli+14c}.  
Moreover, they have been used to unravel the processes that lead to the injection into DSA of protons \citep{caprioli+15}, ions with arbitrary mass/charge ratio \citep{caprioli+17}, and pre-existing CRs \citep{caprioli+18}. 

Here we show some preliminary results obtained with the code \dHybridR, a generalization of the classical hybrid approach that allows accelerated particles to become relativistic \cite{haggerty+19a}.
The reader can refer to the works above for more information about the setup of a non-relativistic shock; the main difference here is that we assumed the fluid electrons to be adiabatic, rather than prescribing an effective polytropic index that mimics ion/electron pressure equipartition in the downstream (see appendix of \cite{caprioli+18}).
We look at the long-term evolution of a quasi-parallel strong shock (with both sonic and Alfv\'enic Mach number $M=20$) to study the non-linear back-reaction of CR acceleration on the shock dynamics. 
The development of a precursor due to the pressure in CRs streaming ahead of the shock was already discussed in \cite{caprioli+14a}, but the focus here is the relation between compression ratio and spectral index. 
For this shock, about 15\% of the shock kinetic energy is converted into accelerated particles \cite{caprioli+14a} and the maximum energy grows linearly in time \cite{caprioli+14c}.

The top panel of Figure \ref{fig:hybrid} shows that the downstream (left in the panel) density increases with time from the canonical value of $R\simeq 4$ for a strong shock to a value $R\gtrsim 7$ when particle acceleration becomes more and more important. 
In \cite{caprioli+14a} such a CR backreaction was partially compensated by the stiffer equation of state for the electrons, which prevented a strong compression of the post-shock plasma.
A total compression ratio $R>4$ is consistent with both the fluid and kinetic predictions of the non-linear DSA theory \cite{jones-ellison91,malkov-drury01} and our simulations represent the first particle-in-cell (PIC) verification of such expectations. 

The next natural step is to compare the  spectral index of the accelerated particles with the standard theory, which at the zeroth order  can be obtained by replacing the test-particle compression ratio $r$ with the total one, $R$,  in Eq.~\ref{eq:qdsa};
this would lead to spectra of relativistic CRs as flat as $q=1.6$ for  $R=6$ ($q=1.1$ for non-relativistic particles, as in our simulation).
In Figure \ref{fig:hybrid}  post-shock ion spectra are compared with the standard DSA prediction (color-coded dashed lines in the middle panel) and with softer power-law distributions (dashed lines in the bottom panel). 
Clearly,  the obtained spectra are indeed much steeper than those predicted by the theory; the best-fitting for their slope can be written as $q_{\rm NLDSA}= q_{\rm DSA}+\frac{3}{R+1}$.

The very nature of such a steepening hinges on the crucial role of the self-generated magnetic turbulence, through a mechanism quite different from those suggested in the literature and outlined in \S\ref{sec:theory};
all the details, along with the analytical formalism, will be discussed in a forthcoming paper (Caprioli, Haggerty \& Blasi, in prog.).
The final goal of our theoretical campaign validated by self-consistent hybrid simulations is to provide the exact amount of steepening as a function of the shock parameters ($\vsh/c$, sonic and Alfv\'enic Mach numbers, inclination) to be compared with the non-thermal phenomenology of diverse space and astrophysical contexts.

\bibliography{Total}
\bibliographystyle{JHEP}

\end{document}